\documentclass[amsmath,amssymb,prb,superscriptaddress, preprint]{revtex4}
\usepackage{graphicx}
\begin{document}
\title{Dimensional crossover of thermal conductance in graphene nanoribbons: A first-principles approach}

\author{Jian Wang}
\email{phcwj@hotmail.com} \affiliation{College of Physical Science and Technology and Center for Complex Science, Yangzhou University, Yangzhou 225002, P. R. China  }

\author{Xiao-Ming Wang}
\affiliation{College of Physical Science and Technology, Yangzhou University, Yangzhou 225002, P. R. China  }

\author{Yun-Fei Chen}
\affiliation{School of Mechanical Engineering, Southeast University, Nanjing, 210096, P. R. China  }

\author{Jian-Sheng Wang}
\affiliation{Department of Physics and Center for Computational Science and Engineering, National University of Singapore, Singapore 117542, Republic of Singapore  }

\date{13 March 2012}

\begin{abstract}
 First-principles density-functional calculations are performed to investigate the thermal transport properties in graphene nanoribbons (GNRs). The dimensional crossover of thermal conductance from one to two dimensions (2D) is clearly demonstrated with increasing ribbon width.  The thermal conductance of GNRs in a few nanometer width already exhibits an approximate low-temperature dependence of $T^{1.5}$, like that of 2D graphene sheet which is attributed to the quadratic nature of dispersion relation for the out-of-plane acoustic phonon modes. Using a zone-folding method, we heuristically  derive the dimensional crossover of thermal conductance with the increase of ribbon width.  Combining our calculations with the experimental phonon mean-free path, some typical values of thermal conductivity at room temperature are estimated for GNRs and for 2D graphene sheet, respectively.  Our findings clarify the issue of low-temperature dependence of thermal transport in GNRs and suggest a calibration range of thermal conductivity  for experimental measurements in graphene-based materials.
\end{abstract}


\maketitle

\section{Introduction}
Since the first exfoliation of graphene,\cite{novoselov2004} many exotic properties have been discovered in  the two-dimensional graphene sheet.\cite{rmprview} Besides graphene, the electronic states of graphene nanoribbons (GNR),\cite{energygap,energygap2,edgegap3,edggeneffct2} thin strips of graphene, can be tuned through
controlling the ribbon width\cite{energygap,energygap2} and the edge chirality.\cite{edgegap3,edggeneffct2} Graphene-based materials can be a technological alternative to silicon semiconductors because of their unique properties.\cite{novoselov2004,rmprview,energygap,energygap2,edgegap3,edggeneffct2} Apart from their electronic properties, graphene turns out to be an excellent heat-conduction material.\cite{balandinreview,balandinaplfirst,balandinnano,lishigroup1,lishigroup2,purdue,geimgroup,cheong,lishigroup0,baowengroup,babarosgroup,damesgroup,twodimsupportedlayer,threedimen, conductivityingnr} However, some discrepant values of thermal conductivity are experimentally reported. Thermal conductivities for the suspended single-layer  graphene flakes  are first measured\cite{balandinaplfirst} using Raman spectroscopy with values in the range from $4.84\pm0.44\times10^{3}$  to $5.30\pm0.48\times10^{3}  \mbox{ W/mK}$ at room temperature.  Using a similar Raman scattering method or utilizing the conventional heat bath method, successive measurements on graphenes \cite{balandinreview,balandinaplfirst,balandinnano,lishigroup1,lishigroup2,purdue,geimgroup,cheong,lishigroup0,baowengroup,babarosgroup,damesgroup,twodimsupportedlayer,threedimen, conductivityingnr} have reported thermal conductivity varying from $\sim170\mbox{ W/mK}$ \cite{babarosgroup} to $\sim5000 \mbox{ W/mK}$\cite{balandinnano,purdue}.  Monolayer of graphene in contact with silicon dioxide has a thermal conductivity of about $600\mbox{ W/mK}$.\cite{twodimsupportedlayer} Room-temperature thermal conductivity\cite{threedimen} decreases from $2800$ to $ 1300 \mbox{ W/mK}$ as the number of atomic planes increases from two to four in few-layer graphenes. Some different theoretical results of thermal conductivity with various approaches have also appeared.\cite{yxu,jwjiang,jswang,ypchen,keblinski,nikatheory,mingo,dimension}

 Furthermore, there is another confusing with regard to the low-temperature dependence of thermal transport in two-dimensional graphenes.  For one-dimensional quantum atomic chains,\cite{onedime} thermal conductance  at low temperatures is proportional to temperature $T$ with the quantized universal coefficient $\pi^2k_{B}^{2}/3h$, where $k_B$ is the Boltzmann constant and $h$ is the Planck constant. For two-dimensional materials, it is well known that thermal conductance depends on temperature as Debye $T^{2}$ law.  A calculation by the valence-force field demonstrates that thermal conductivities at low temperature conform\cite{nikatheory} to the conventional relation of $T^{2}$. But a theoretical estimation\cite{mingo} and the continuum model\cite{dimension} propose a  $T^{1.5}$ temperature dependence of thermal conductivities for graphene, followed by some recent experimental observations.\cite{lishigroup0,baowengroup,babarosgroup,damesgroup}

  To understand thermal transport in graphene-based  materials is not only important to the possible technological applications but also crucial for a fundamental understanding of thermal properties in low-dimensional systems. The reduced-dimension effects on thermal transport are becoming critical to both device reliability and intrinsic physics. Temperature dependence of thermal transport, especially in the low-temperature range, is essential for understanding the fundamental physics behind the measured thermal conductivities.  An accurate knowledge of the temperature dependence of thermal transport is important because it helps to identify phonon branches which have the dominant contributions to heat conduction.  Low-frequency acoustic phonon branches which determines the behavior of low-temperature thermal transport are sensitive to the weak-bonding interactions between the long-range atoms. The valence-force field considering only the nearest neighbors or the continuum model may not be an appropriate model. Moreover, the experimental measurements of temperature dependence alone cannot provide evidence in favor of one or the other phonon contribution because the temperature dependence in graphite depend strongly on the material quality.\cite{balandinreview} In contrast, the first-principles approach can yield accurate phonon dispersion\cite{phonondisp} relations without empirical parameters, including the long-range interactions among the atoms.

  In this paper, we report a systematic investigation on temperature dependence of thermal transport in graphenes varying from quasi one-dimensional nanoribbons to infinite two-dimensional sheets via first-principles density-functional theory calculations.\cite{siesta}  The dimensional crossover of thermal conductance from one to two dimensions is clearly observed in GNRs with increasing ribbon width. We find that the thermal conductance of GNRs of a few nanometers wide already exhibits an approximate low-temperature dependence of $T^{1.5}$. Ballistic thermal conductance of wide ribbons is also found to converge towards the corresponding values of 2-D graphene sheet at low temperature. Combining our calculations with the experimental phonon mean-free path(MFP),\cite{balandinaplfirst} typical values of thermal conductivity are estimated at room temperature for GNRs and 2D graphene sheet, respectively. We conclusively clarify the issue of low-temperature dependence of thermal conductance for graphene-based materials and suggest a calibration range of thermal conductivity  for experimental measurements. We expect that our first-principles results may provide some valuable insights into the experimental measurement of thermal transport in graphene-based materials.

\section{Computational Methods and Formulas}
 To calculate the thermal conductance, we should obtain force constants in graphene nanoribbons and in 2D graphene sheet. The nanoribbons are classified\cite{phonondisp} by the number of dimers in unit cell using a common convention, where armchair-edged nanoribbons or zigzag-edged nanoribbons with $n$ dimers are referred to as \textbf{n}-AGNR or \textbf{n}-ZGNR. To eliminate the dangling bonds at the edge of ribbon, we have passivated the nanoribbon edges with hydrogen atoms.  The interatomic force constants are obtained from a series of supercell calculations using the software packages SIESTA\cite{siesta} based on the first-principles density-functional approach. The ribbons are periodic with a supercell of 5 unit cells along the ribbon axis for minimizing the periodic continuation.  In other two directions, the periodic image of ribbon is separated from its nearest neighbors at least by 18 $\mbox{\AA}$   to prevent interactions between them. We use a $7\times7\times1$ supercell for calculations of the graphene sheet, with a space 20 $\mbox{\AA}$  between periodic images of sheet.  For DFT calculations, Troullier-Martins pseudopotentials are used for both C and H atoms. The valence electrons are described by a $\mathrm{double}-\zeta$ basis set plus polarization orbits (DZP). The local-density approximation (LDA) with the exchange-correlation functional due to Ceperley-Alder is employed. To suppress the force fluctuations, we use a rather fine Mesh cutoff of 400 Ry during calculations.

 We first perform geometrical optimization with a force tolerance 0.001 $\mathrm{eV}/\mbox{\AA}$ with the aforementioned parameters  using a conjugated-gradient minimization method. Force constants then are derived from the optimized structures with the method of finite difference, where each atom is displaced from its equilibrium position by a distance of $\pm0.0212\mbox{\AA}$. After forces are evaluated, a central finite difference with respect to the displacement is utilized to compute the fore constants. We have calculated force constants for the \textbf{n}-AGNRs, \textbf{n}-ZGNRs with $n=2,3,\cdots,14$ dimers in width and for the two-dimensional graphene sheet, respectively.

 Thermal conductivity $\kappa$ along one direction, e.g. the $z\mathrm{-axis}$ direction, can be expressed by the well-known Boltzmann-Peierls formula\cite{callaway}
 \begin{equation}
\label{thermal_conductivity} \kappa_z =\sum_{p,v_{n}^{z}>0}\int \frac{d^{n}\bf q}{(2\pi)^{n}}\hbar\omega_{p}\frac{\partial{f_B}}{\partial{T}}v_{p}^{z}l_{p},
\end{equation}
 where $p$ denotes the polarization of phonon, $\bf{q}$  wave vector, $\omega_p$ the angular frequency, $f_B$ the Bose-Einstein distribution at temperature $T$. The phonon group velocity $v_{p}^{z}$ along the transport direction $z$ is given by $v_{p}^{z}=\partial \omega_p/\partial q_z$ and $l_p$ represents the length of phonon mean-free path (MFP). The number $n=1,2 \mbox{ and }3$ denotes the system dimension. The integration over the wave vector is carried out within the first Brillouin zone. Graphene has a long phonon MFP, which is estimated to be about $775$ nm near room temperature.\cite{balandinaplfirst} Therefore, at low temperatures, it is reasonably expected that phonon transport in GNRs or in 2-D graphene sheets will be dominantly ballistic or quasi-ballistic rather than diffusive. In regime of ballistic or quasi-ballistic transport, thermal conductivity is not well defined because thermal conductivity $\kappa$ depends on the length of system. Assuming that the MFP $l_p$ for ballistic or quasi-ballistic transport is independent of the polarization since it can be compared with the length of system in experiments,  we thus rewrite Eq.~(\ref{thermal_conductivity}) using  the mean value theorem  as
 \begin{subequations}
\label{thermal_conductance}
\begin{eqnarray}
\kappa_z &= & \sigma \bar{l},  \\
\mathrm{with} \quad \sigma&\equiv&\sum_{p,v_{p}^{z}>0}\int \frac{d^{n}\bf q}{(2\pi)^{n}}\hbar\omega_{p}\frac{\partial{f_B}}{\partial{T}}v_{p}^{z}.
\end{eqnarray}
\end{subequations}
Here $\bar{l}$ is the average length of phonon MFP, $p$ denotes the polarization and $\sigma$ is defined as thermal conductance per unit area. In particular, thermal conductance $\sigma$ in Eq.~(\ref{thermal_conductance}) for  one-dimensional ballistic transport is reduced to the following expression
\begin{equation}
\label{one_dimensional_thermal_conductance}
\sigma_{1D}=\frac{1}{S}\sum_{p,v_{p}>0}\int \frac{d \mbox{q}}{2\pi}\hbar\omega_{p}\frac{\partial{f_B}}{\partial{T}}v_{p},
\end{equation}
where $S=h\cdot w$ is the cross-section, $h$ the thickness and $w$ the width. In contrast with thermal conductivity, thermal conductance is uniquely defined in the regime of ballistic transport.  Here we concentrate on the salient features of temperature dependence of thermal transport at low-temperature in both GNRs and graphenes such that nonlinear scattering characterized by MFP is first neglected. Phonon transmission is assumed to be ballistic such that thermal transport at low temperature is distinguished by thermal conductance $\sigma$ as defined in Eq.~(\ref{thermal_conductance}). This is a valid approximation for thermal transport in graphenes at low temperatures, where the length of MFP is approaching the system size.  Nonlinear scattering is later included through the average length of phonon MFP in the end of paper.

\section{Results and Discussion}
\subsection{Phonon dispersion relations}
\begin{figure}[hb]
\includegraphics[width=0.8\columnwidth]{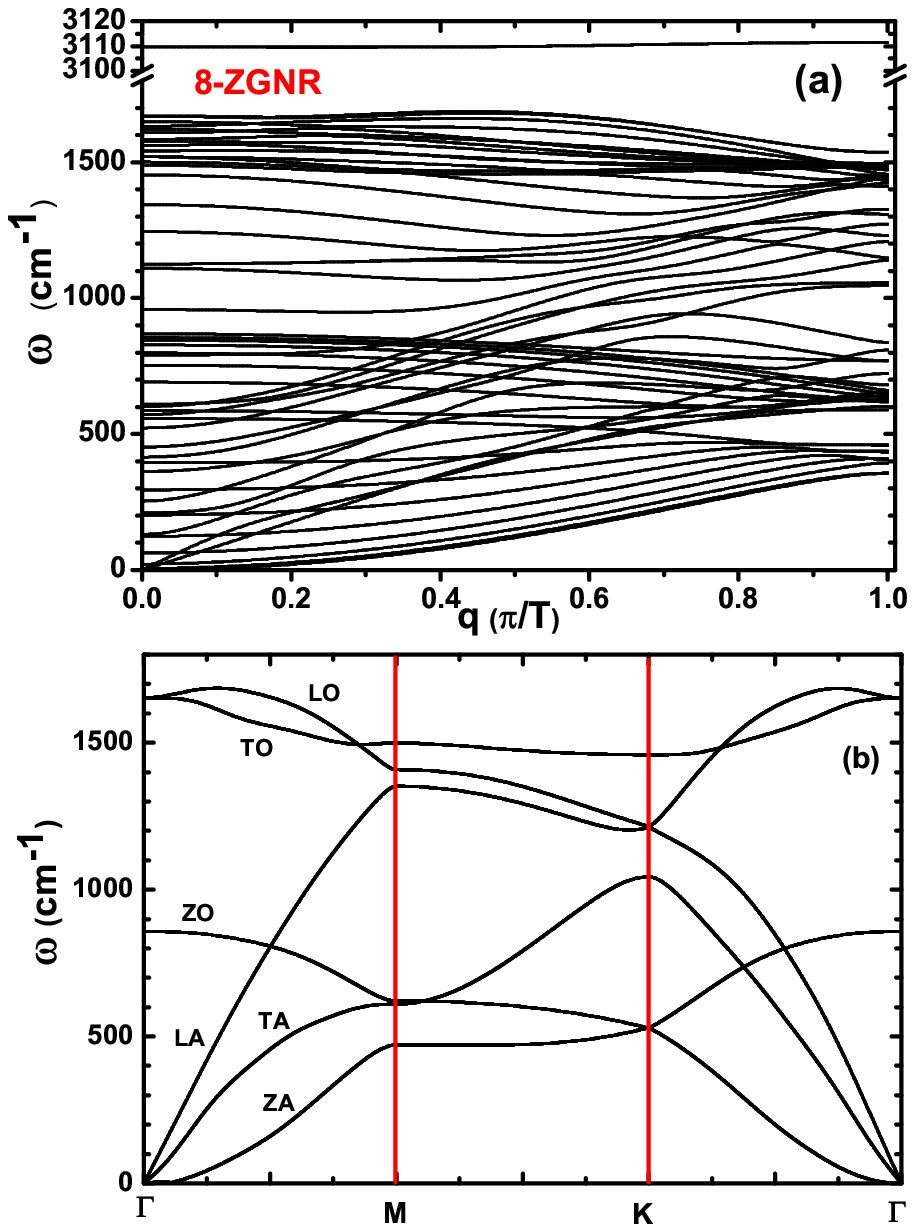}
\caption{\label{fig1:phonondisperion} (Color online).  Phonon dispersion relations calculated using first-principles density-functional calculations. (a) The dispersion relation for an 8-ZGNR. The wave number $q$ is given in units of the length of unit cell $T$. (c) The phonon dispersion curves  plotted  along high symmetry directions for a two-dimensional graphene sheet.  }
\end{figure}

Since thermal conductance is intrinsically related to phonon dispersion relations, we first present the calculated phonon dispersions from first principles.  Fig.~\ref{fig1:phonondisperion} shows the calculated phonon dispersion for  the 8-ZGNR and 2-D graphene sheet.  It can be seen from the figure that the stretching modes due to the C-H bonding in GNRs are located about $\mbox{3110 cm}^{-1}$. We can also find that the highest frequency of longitudinal-optical(LO) mode at $\Gamma$ point is about $\mbox{1650 cm}^{-1}$ for both GNRs and the two-dimensional graphene sheet. For GNRs, there are four acoustic phonon branches near $\Gamma$ point, including an out-of plane mode(ZA), an in-plane transverse mode(TA), an in-plane longitudinal(LA) and a torsion branch that will disappear with the increasing width. In comparison, similar acoustic branches except the torsion one are obtained for the 2-D graphene sheet. As shown in Fig.~\ref{fig1:phonondisperion}, the ZA mode shows a quadratic energy dispersion  around the $\Gamma$-point as a consequence of the point-group symmetry of graphene\cite{nanotubebook1} while the TA and LA modes display a linear dispersion.  The phonon dispersion curves both for the GNRs and 2-D graphene sheet are consistent with the previous reported results.\cite{phonondisp}

The phonon dispersion  between the GNRs and the 2-D graphene sheet can be mutually transformed through the unfolding of nanoribbons's Brillouin zone to that of graphene.\cite{phonondisp} The phonon dispersions of the 2-D graphene sheet comprise three acoustic branches and three optical branches as indicated in Fig.~\ref{fig1:phonondisperion}(b) because there are two carbon atoms in the unit cell of graphene. In GNRs, a group of six modes can be found  equivalent to the six phonon modes of graphene, with respect to the phonon eigenvectors near $\Gamma-\mbox{point}$.  Hence, all phonon modes of GNRs can be interpreted as these six fundamental modes and their overtones.\cite{phonondisp}

\subsection{Thermal conductance of graphene nanoribbons}
\begin{figure}[ht]
\includegraphics[width=0.9 \columnwidth]{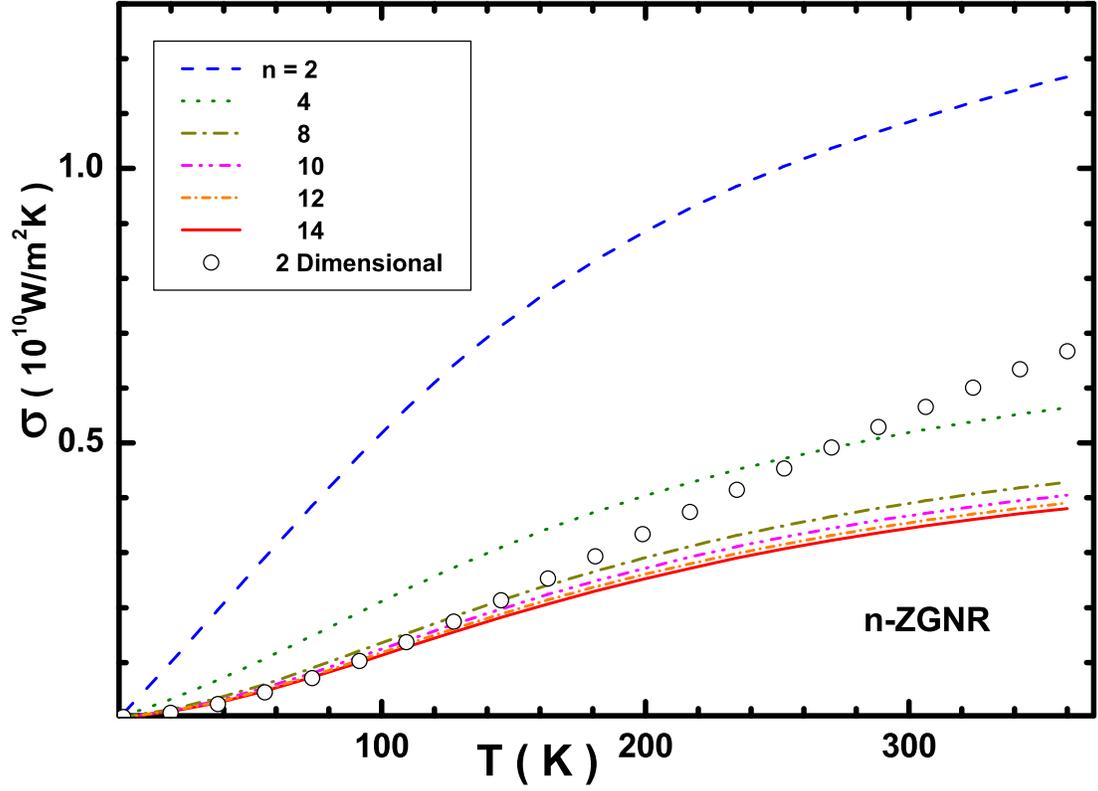}
\caption{\label{fig2:thermalconductance} (Color online).  Temperature dependence of thermal conductance per unit area for ZGNRs with the increase of ribbon width $n$, which indicates the number of dimers per unit cell. The open circles in the figure represent thermal conductance along $\Gamma-K$ direction, corresponding to the axis direction of ZGNR.   }
\end{figure}

Using the complete phonon dispersion relations, we have systematically calculated thermal conductance $\sigma$ per unit area of Eq.~(\ref{thermal_conductance}) for GNRs with increasing width and for the 2-D graphene sheet, as shown in Fig.~\ref{fig2:thermalconductance}. During the calculation, the layer thickness $h$ for both GNRs and the graphene sheet  is taken a typical value $h=0.335\mbox{ nm}$.  The ribbon width $w$ for ZGNRs and AGNRs is computed by the formula $w_{{\rm ZGNR}}=\sqrt{3}(n-1)a_0/2$ and $w_{{\rm AGNR}}=(n-1)a_0/2$, respectively. Here $n$ is the number of dimers in unit cell and $a_0$ denotes the graphene lattice constant given by $a_0=0.246\mbox{ nm}$. For simplicity, a very small deviation of ribbon width after the structure relaxation is not considered in the formula. Fig.~\ref{fig2:thermalconductance} shows  thermal conductance per unit area as a function of temperature for the n-ZGNRs with the increasing number of dimers $n$.  The open circles in Fig.~\ref{fig2:thermalconductance} represent thermal conductance along $\Gamma-K$ direction, corresponding to the axis direction of ZGNR. For a clear display of low-temperature dependence of thermal conductance, a log-log plot of thermal conductance for $\mbox{T}<200\mbox{K}$  is further illustrated in Fig.~\ref{fig3:dimesional}.

\subsubsection{Dimensional crossover of thermal conductance }
\begin{figure}[hb]
\includegraphics[width=0.9\columnwidth]{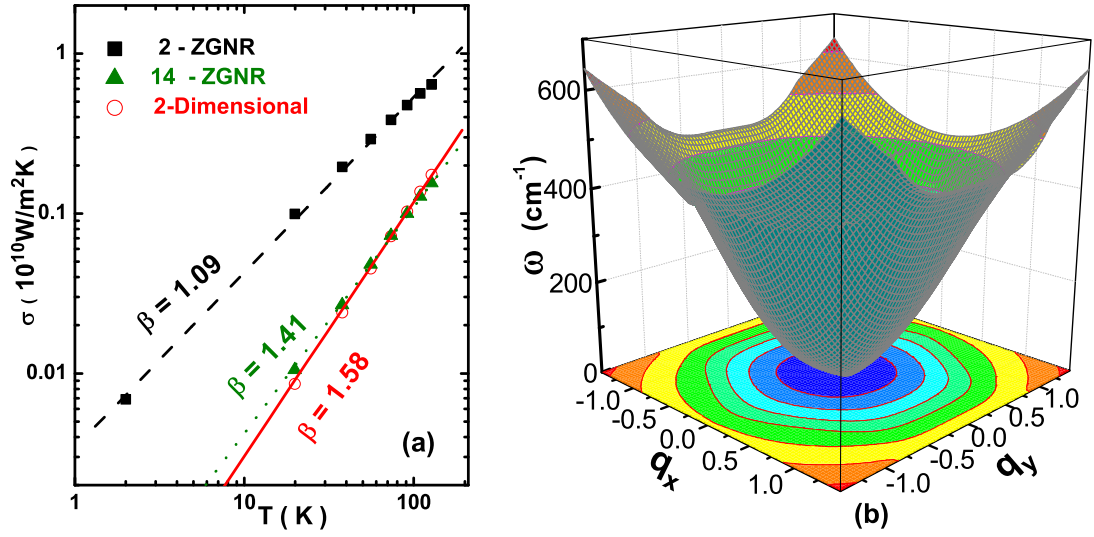}
\caption{\label{fig3:dimesional} (Color online). (a) The scaling behavior of thermal conductance with temperature for the 2-ZGNR, 14-ZGNR and two-dimensional graphene sheet, respectively. The dashed, dotted, and solid lines in the figure are numerically fitted as $\mathrm{T}^\beta$ at low temperatures. (b) The 3D-view and its projection of phonon dispersion relations for the out-of-plane acoustic branch(ZA) in the first brillouin zone. }
\end{figure}

 The most significant feature of thermal conductance is the scaling behavior in the low-temperature dependence for ZGNRs.  To quantitatively show such temperature dependence, the curves of thermal conductance $\sigma$ fitted vs temperature as $\mbox{T}^{\beta}$ below $200\mbox{ K}$ are also plotted in Fig.~\ref{fig3:dimesional}.  We can find that the exponent $\beta$ increases from $\beta=1.09$ to $\beta=1.41$, approaching to the value of the two-dimensional graphene sheet $\beta=1.58$, when the number of dimers for ZGNRs grows from $n=2$ to $n=14$. This dimensional crossover of thermal conductance from one to two dimensions is clearly demonstrated in Fig.~\ref{fig3:dimesional}(a). It is well known that thermal conductance at low temperatures is proportional to temperature $\mathrm{T}$ for one-dimensional quantum atomic chains\cite{onedime} whereas the low-temperature dependence of $\mbox{T}^2$  for the two-dimensional phonon gas is described by the conventional Debye law. For 2D graphene sheet, a calculation by the valence-force field demonstrates that thermal conductivity at low temperature conform\cite{nikatheory} to the conventional relation of $\mathrm{T}^{2}$.  But a theoretical estimation\cite{mingo} propose a  $\mathrm{T}^{1.5}$ temperature dependence of thermal conductance for an infinite 2D graphene sheet. An analysis of vibrational modes\cite{dimension} using the continuum mechanics shows that GNRs with width $w>500\mbox{ nm}$ will show a low-temperature dependence of $\mbox{T}^{1.5}$. In comparison with the approaches of the empirical valence-force and continuum mechanics, the density-functional theory can yield an accurate phonon dispersion relations without empirical parameters.  Our first-principles calculations shown in Fig.~\ref{fig3:dimesional}(a) clarify that the two-dimensional graphene sheet will display a low-temperature dependence of $\mbox{T}^{1.5}$ and that dimensional crossover of thermal conductance from one to two dimensions will be expected for the GNRs of only a few nanometers in width. We conclude that thermal conductance for most samples of graphene flakes in experiments  with a width beyond a few nanometers will be characterized by a low-temperature dependence of $\mbox{T}^{1.5}$.  Our calculated results are consistent with the experimental results,\cite{lishigroup0,baowengroup,babarosgroup,damesgroup}  where the width of graphene samples is on the order of micrometers.

Next we elucidate the reasons for the dimensional crossover of thermal conductance in GNRs through a heuristic derivation. The phonon wave vector $\mbox{\bf q} \equiv (q_t, q_l)$ for GNRs consists of the transverse direction component $q_{t}$ perpendicular to the ribbon axis and the longitudinal component $q_l$ along the ribbon axis. The longitudinal wave vector $q_l$ represented by the x-axis in Fig.~\ref{fig1:phonondisperion}(a) is continuous because the ribbon length goes to infinite. But the transverse wave vector $q_{t}$ is discrete, due to the finite ribbon width which only allows the standing waves with the boundary condition $q_t w= k\cdot\pi,\quad k=0,1,\cdots,n-1$. Here $w$ denotes the ribbon width and $n$ is the number of dimers in width. Therefore, the transverse wave vector $q_t$ can only take the quantized values given by $q_t=k\cdot\pi/w$. Each of the discrete value of $q_t$ corresponds to one branch of over-tone phonon mode.\cite{phonondisp} The vibrational modes in GNRs  shown in Fig.~\ref{fig1:phonondisperion}(a) can be classified\cite{phonondisp} into: six fundamental modes equivalent to the phonon modes in 2-D graphene sheet, over-tone modes of the fundamental modes and the C-H modes resulting from the passivation with hydrogen at edges. The contribution to thermal conductance from the C-H modes can be neglected at low temperature on account of their high frequencies. Hence, the summation $p$ in Eq.~(\ref{one_dimensional_thermal_conductance}) over different polarization branches can be separated into the terms of six basic modes plus their over-tone modes, expressed by $\Sigma_{p}\rightarrow\Sigma_{p'}\Sigma_{k=0,n-1}$. Here $p'$ denotes the summation over six fundamental phonon modes. When the number of dimers $n$ in unit cell is large enough, we can transform the discrete summation $k$ into the integration over continuum $q_t$, mathematically described by $\frac{2\pi}{w}\Sigma_{k=0,n-1}\rightarrow\int d\mbox{q}_{t}$. In other words, for the nanoribbon with enough width, thermal conductance for one-dimensional transport in Eq.~(\ref{one_dimensional_thermal_conductance}) will transition into the two-dimensional expression $\sigma_{2D}=\frac{1}{h}\Sigma_{p'}\frac{1}{(2\pi)^2}\int d\mbox{q}_td\mbox{q}_l\hbar\omega_{p'}\frac{\partial{f_B}}{\partial{T}}v_{p'}^{z}$ with the constraint of $v_{p'}^{z}>0$, where $h$ is the thickness. Here $p'$ denotes the basic phonon mode, ranging from LA,TA,and ZA to LO,TO and ZO modes shown in Fig.~\ref{fig1:phonondisperion}(b) for the 2-D graphene sheet. From our numerical calculations, we find that the number of dimers $n=14$ already leads to low-temperature thermal conductance fairly close to that of 2-D graphene, demonstrating a dimensional crossover from one to two dimensions.

\subsubsection{Temperature dependence of thermal conductance }
Further we turn to the causes of the low-temperature dependence of $T^{1.5}$ in 2-D graphene sheet through an analytic derivation. The key reason lies in the different contributions to thermal conductance made by each of the polarized phonon branch with different dispersion relations. Note that we only need to consider the acoustic branches during calculating thermal conductance at low temperature. It can be observed from Fig.~\ref{fig1:phonondisperion}(b)  that both LA and TA branches exhibit the linear phonon dispersion relations at low-frequencies while the ZA branch display an approximately quadratic dispersion curve near $\Gamma$ point. To illustrate the dispersion relations in full phase space, we have presented the three-dimensional view  of the ZA branch of the first brillouin zone in Fig.~\ref{fig3:dimesional}(b). A direct proof of contribution to thermal conductance proceed as follows. In the integration of Eq.~(\ref{thermal_conductance}), we can take the integration over frequency $\omega$ in substitute of the wave vector $\bf q$ by introducing the delta function,
\begin{subequations}
\label{thermal_conductance2d}
\begin{eqnarray}
\sigma_{2D} &\propto & \sum_{p}\int\frac{d^2\mbox{\bf q}}{(2\pi)^2}\hbar\omega_p\frac{\partial{f_B}}{\partial{T}}v_{p}^{z},  \\
\quad & = & \frac{1}{(2\pi)^2}\int d\omega\hbar\omega\frac{\partial{f_B}}{\partial{T}}\mathcal{T}[\omega],\\
\mathrm{with}\quad \mathcal{T}[\omega] &\equiv &\sum_{p}\int d^2\mbox{\bf q}\delta(\omega-\omega_p)v^z_{p}.
\end{eqnarray}
\end{subequations}
Here $\delta$ is the dirac delta function.  $\mathcal{T}[\omega]$ is defined as the effect transmission,  which means the number of phonon branches at a given frequency $\omega$ for one-dimensional transport. For two-dimensional transport, it is straightforward to prove that $\mathcal{T}[\omega]$ behaves as  $\omega^{1/2}$ for quadratic phonon dispersion relation while $\mathcal{T}[\omega]$ is proportional to $\omega$ with regard to linear phonon dispersion relation. Assuming that $\mathcal{T}[\omega]\propto\omega^{\alpha}$, it can be easily derived that thermal conductance has low-temperature dependence as $\sigma_{2D}\propto T^{1+\alpha}$. Therefore, the contribution to thermal conductance  made by the ZA branch with quadratic phonon dispersion scales as $T^{1.5}$  while thermal conductance resulting from the linear LA and TA branch increases with temperature as $T^{2}$.

The reason why  whole thermal conductance at low temperature increases with temperature as $T^{1.5}$ like that of the ZA branch roots in differences of the vibrational density of states(vDOS), which can be defined as $\mbox{vDOS}[\omega]\propto\int d^2\mbox{\bf q}\delta(\omega-\omega_p)$.  It can be easily verified that $\mbox{vDOS}$ in two dimensions is proportional to $\omega$ for the phonon branch with linear dispersion while two-dimensional $\mbox{vDOS}$ for the quadratic phonon branch remains to be constant. Accordingly, as the phonon frequency tends to zero,  the number of vibrational modes expressed by vDOS  also decreases to zero for branches with the linear phonon dispersion. On the other hand, the number of vibrational modes for the quadratic phonon branch does not change with frequencies.  At low temperature, only low-frequency phonons mainly contribute to thermal conductance. Therefore, the ZA phonon branch with the quadratic dispersion makes a dominant contribution to low-temperature thermal conductance due to its constant vibrational density of states. By contrast, the contributions resulting from the linear LA and TA phonon branches are trivial because of their diminishing vibrational modes at low frequencies.  Namely, thermal conductance at low temperature for two-dimensional graphene sheet increases with temperature as $T^{1.5}$ on account of the dominant contributions made by the ZA phonon branch.

\subsubsection{Convergence of low-temperature thermal conductance and anisotropy.  }
\begin{figure}[hb]
\includegraphics[width=0.9\columnwidth]{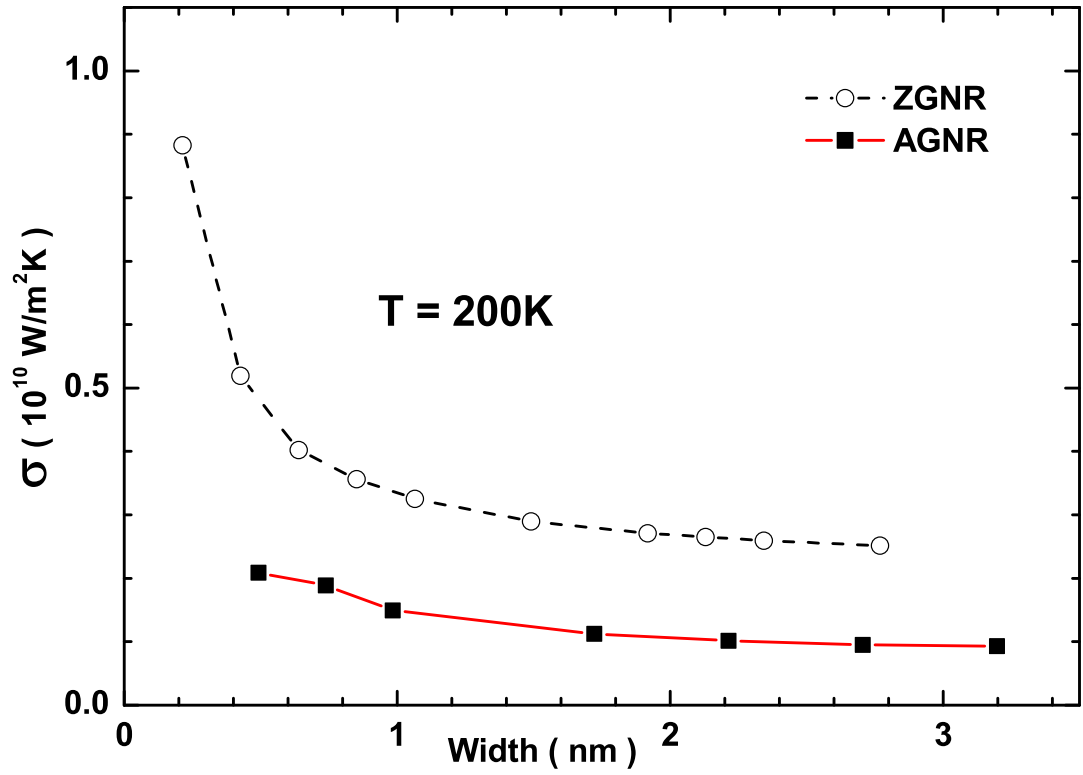}
\caption{\label{fig4:width} (Color online).  Thermal conductance per unit area as a function of the ribbon width for the zigzag and armchair graphene nanoribbons at temperature $T=200K$.   }
\end{figure}

We now discuss another interesting feature observed in Fig.~\ref{fig2:thermalconductance}. It can be seen from the figure that thermal conductance of GNRs at low temperatures converges to that of 2D graphene sheet with the increase of ribbon width. A systematical investigation on thermal conductance as a function of the ribbon width for the zigzag and armchair graphene nanoribbons at temperature T=200K is shown in Fig.~\ref{fig4:width}. We can find that thermal conductance per unit area both for ZGNRs and AGNRs decreases with the increase of ribbon width, converging  to a stable value. At low temperatures, the value of convergence agrees well with that of 2D graphene sheet along the corresponding direction, as shown by the open circles in Fig.~\ref{fig2:thermalconductance}. Such agreement can be explained by the fact that with increasing width phonon dispersion relations of GNRs, especially at low frequencies, are approaching that of 2D graphene sheet as described in the foregoing paragraphs. By contrast, there are some value differences in thermal conductance at high temperatures between GNRs and 2D graphene. The deviation results from the scattering caused by the confinement from the edges. Because of the confinement of standing wave due to the ribbon edges, the phonon dispersion curves tend to be flat at high frequencies, yielding a small group velocity. Thus, the reduced group velocity for high-frequency phonon branches in GNRs leads to smaller thermal conductance at high temperature. In addition, Fig.~\ref{fig4:width} shows that thermal conductance per unit area of the ZGNRs is larger than that of AGNRs. This behavior originates from the anisotropy of phonon dispersion relationships in consistency with  other simulation results.\cite{yxu, jwjiang, jswang, ypchen}

\subsection{Nonlinear scattering and thermal conductivity}
Finally, we include the nonlinear phonon scattering effects through the phonon MFP. The above calculations have not taken into account phonon-phonon scattering. At high temperatures, the nonlinear phonon scattering is inevitable for more realistic situations. However, to calculate nonlinear phonon scattering for GNR system with  a few hundred atoms is a formidable task for the firs-principles approach and is also beyond the ability of any first-principles implementations. For simplicity, we phenomenologically introduce the phonon-phonon scattering into the present ballistic theory through the phonon MFP described by Eq.~(\ref{thermal_conductance}). It is estimated from experiments that the phonon MFP at room temperature\cite{balandinaplfirst} on average is about 775 nm  for graphene samples.  At T=300K, our calculated thermal conductance per unit area for 2D graphene sheet is $\sigma_{2D}=5.5\times 10^9 \mbox{ W/}\mbox{m}^2K$  along $\Gamma-K$ direction and the converged value of $\sigma_{ZGNR}=3.4\times 10^9 \mbox{ W/}\mbox{m}^2K$ is obtained for the ZGNRs as shown in Fig.~\ref{fig2:thermalconductance}. In comparison with the previous estimation of  thermal conductance\cite{mingo} for 2D graphene at room temperature $\sigma_{2D}=3.1\times 10^9 \mbox{ W/}\mbox{m}^2K$ given by the formula $\sigma=0.6\times10^6T^{3/2} {\rm W/m^2K^{5/2}}$, our results are qualitatively consistent with this estimation. Combining the calculated thermal conductance with the experimental phonon MFP, thermal conductivity at room temperature $\kappa_{2D}=4263\mbox{ W/mk}$ for 2D graphene and $\kappa_{GNR}=2635 \mbox{ W/mk}$ for ZGNRs is achieved. Even though our calculated results cannot be directly  compared with experimental results where some other factors exist, such as structure defects, supporting substrate and surface tensions, etc.,  a calibration range of thermal conductivity in GNRs and two-dimensional graphene is suggested for experimental measurements from our calculations. We propose that the room-temperature thermal conductivity of GNRs may be estimated on the order of $\sim2600\mbox{ W/mK}$ for GNRs and of $\sim4200\mbox{ W/mK}$ for 2D graphene sheet if  intrinsic phonon-phonon scattering is only considered. We think that the measurement of intrinsic thermal conductivity in GNRs or in 2D graphene sheet can be reduced to measuring the accurate phonon mean-free path.\cite{chen}

\section{Summary and Conclusions}
In summary, we have investigated low-temperature thermal transport in GNRs and 2D graphene sheet using first-principles density-functional theory approach. A dimensional crossover of thermal conductance from one dimensional GNRs to two dimensional graphene sheet is clearly demonstrated. We determine that thermal conductance of GNRs with a width of a few nanometers will exhibit a low-temperature dependence of $T^{1.5}$. A heuristical derivation of thermal conductance by the zone-folding method is carried out to elucidate the reason for this dimensional crossover. The reason for the  temperature  dependence of $T^{1.5}$  can be attributed to the quadratic nature of dispersion relation for the out-of-plane acoustic phonon branch.  In addition, we find that low-temperature thermal conductance in GNRs converges towards the corresponding values of 2D graphene sheet with the increase of ribbon width. Combining our calculations with the experimental phonon MFP, typical values of thermal conductivity for GNRs and for 2D graphene sheet are estimated at room temperature. Our findings conclusively clarify the issue of the low-temperature dependence of thermal conductance for GNRs and suggest a calibration value range of thermal conductivity in graphene-based material. We hope that our findings can offer some enlightening insights into the experimental measurement of thermal transport in graphene-based materials.

\section{Acknowledgement}
J.W. acknowledges the support from National Natural Science
Foundation of China (NSFC) under the grant 10705023 and 11075136, as
well as from Jiangsu Natural Science Foundation under the grant
BK2009180. J.-S. W. is supported by a URC grant R-144-000-257-112.

\newpage

\end{document}